\documentclass[aps,prd,twocolumn,flushleft,amsmath,amssymb,nofootinbib]{revtex4}
\usepackage{subfig}
\usepackage{graphicx}
\usepackage{dcolumn} 
\usepackage{bm}      

\usepackage{hyperref} 
\usepackage{amsmath}

\newcommand{\bq}{\begin{equation}}
\newcommand{\eq}{\end{equation}}
\newcommand{\ba}{\begin{eqnarray}}
\newcommand{\ea}{\end{eqnarray}}

\begin{document}

\title{Emergence of scalar matter from spinfoam model}

\author{Peng Xu\footnote{xupeng@mail.bnu.edu.cn} and Yongge Ma\footnote{mayg@bnu.edu.cn}}
\affiliation{Department of Physics, Beijing Normal University, Beijing 100875, China}

\begin{abstract}
A spinfoam model of 3D gravity non-minimally coupled with a scalar
field is studied. By discretization of
the scalar field, the model is worked out precisely in a
purely combinational way. It is shown that the quantum physics of
the scalar matter are totally encoded into the modified dynamics of
$SU(2)$ spin-network states which describe the quantum geometry of space.
It turns out that the physics of the scalar matter coupled with gravity manifested in the low energy
scale can be viewed as the phenomena emerged from this microscopical
construction. This gives rise to a radical
observation on the issue of the unification of geometry and
matter.
\end{abstract}

 \maketitle

The search for the mergence of quantum mechanics and general
relativity has a long history that can be traced back to the time of
 Einstein and Dirac. Over the past twenty years the so-called loop
quantum gravity (LQG) has made considerable progress in quantizing
general relativity background independently. While the
 \emph{kinematics} of LQG is well established, the \emph{dynamics} of the theory is still
 a thorny problem \cite{AshtekarReview,RovelliBook,ThiemannBook,HanReview}.
Spinfoam models were first introduced as candidates to solve
 this problem \cite{ReisenbergerSumSyrface,RovelliProjector}. Remarkably, a large number
 of distinct approaches to the issue of quantum gravity converge to this
formalism\cite{PerezReview,OritiThesis}. Spinfoam models can be
interpreted as the Feynman sum over histories of the evolutions of
quantum geometries. Penrose's idea, that quantum spacetime should be
described and controlled in a purely combinational way by means of
couplings of angular momentums \cite{Penrose}, was somehow
re-emerged in LQG by the construction of spin-network states
describing the quantum geometry \cite{SpinNW}.

 A specified
spinfoam is a two complex $\Gamma$ constructed from the dual of
the triangulation $\triangle$ of the spacetime manifold with its faces
colored by spins $j_{f}$  and edges colored by intertwinors
$\iota_{e}$, which encode the geometric data of the simplicial
manifold. The dynamics is then determined by the sum over amplitudes
 contributed by all possible spinfoams
\bq
\mathcal{Z}=\sum_{\Gamma}\mathbf{w}(\Gamma)\sum_{j_{f},\iota_{e}}
\prod_{f\in\Gamma}\mathcal{A}_{f}(j_{f})\prod_{e\in\Gamma}
\mathcal{A}_e(j_f,\iota_{e})\prod_{v\in\Gamma}\mathcal{A}_{v}(j_{f},\iota_{e}),\nonumber{}
\eq
where $\mathbf{w}(\Gamma)$ denotes the weight associated to each
triangulation, $\mathcal{A}_f$, $\mathcal{A}_e$
and $\mathcal{A}_{v}$ are the amplitudes associated
to each face, edge and vertex of $\Gamma$ respectively. The choices
of different functions $\mathbf{w}(\Gamma)$ and the amplitudes
$\mathcal{A}_f$, $\mathcal{A}_e$, and $\mathcal{A}_{v}$ define
different models. This picture can be viewed as a 2D generalization
of Feynman diagrams, and the vertex amplitude plays the similar role
as in standard QFT. In recent years the most active research fields
of spinfoam models are the analysis of their low energy
limits\cite{RovelliScatter,RovelliPropagator}, their connection with
canonical programs\cite{PerezConnectLQG,Projector2008}, and the
matter couplings. The problem of matter couplings in a quantum
gravity theory is of extremely importance, because it is necessary
to understand how matter fields interact with gravity in a fully
quantum mechanical way. Various methods have been proposed to solve
this problem in the formalism of spinfoams. One of them is to
incorporate the Feynman diagrams of matter interactions into the
spinfoams\cite{PerezParticle,FreidelRegge,FreidelGFT,OritiGFT}. A
surprising result is that the effective dynamics of a scalar field
coupled to 3d spinfoams can be described by a non-commutative field
theory, which encodes the information of quantum
spacetime\cite{FreidelNonCommu}.  Fermions and gauge fields coupled to 3D spinfoam
are studied in \cite{FairbairnFermion,SpezialeGauge}. Another more ambitious approach to
this issue was proposed in \cite{SmolinBraid} \cite{WanBraid}, where
particles of standard model are suggested as the local excitations
of the quantum states of spacetime. In this letter we propose a new
approach to incorporate a scalar field $\phi$ into 3D spinfoams.
Here the dynamics is determined by a new vertex amplitude
$\mathcal{A}_v(j_f,\iota_e,\Phi_e)$ which involves the degrees of
freedom of both the geometry and the scalar matter. The attractive
property of this new model is that it can be casted into a modified
dynamics of pure quantum geometry. Thus the scalar matter manifested
itself in the low energy scale can be viewed as being emerged from
this microscopical construction.

We start with the physical system of a massless Klein-Gordon field
$\phi$ coupled to the gravitational field with Riemannian signature defined
on a 3D manifold $\mathcal{M}$. The standard action reads
\bq
S[g,\phi]=\int_{\mathcal{M}}\sqrt{|g|}(R[g]-g^{ab}\nabla_{a}\phi\nabla_{b}\phi).
\eq It can be shown that the classical dynamics of this action is
``conformally'' equivalent to that of \bq
S[\tilde{g},\tilde{\phi}]=\int_{\mathcal{M}}\sqrt{|\tilde{\phi}
|}\sqrt{|\tilde{g}|}R[\tilde{g}] \label{action1},
\eq
under the transformations
$g_{ab}=|\tilde{\phi}| \tilde{g}_{ab}$ and $\phi=-\frac{\sqrt{2}}{2}ln|\tilde{\phi}|$ \cite{MaKilling4D}.
 The action (\ref{action1}) is also of
physical interest because it can be obtained by the symmetric reduction
from a 4D spacetime with a hypersurface orthogonal Killing vector
field\cite{MaKilling4D}. In the first order
formalism this action is written in terms of $(e^I_a,\omega^{IJ}_a)$ as
\bq
S[e,\omega,\phi]=\int_{\mathcal{M}}\sqrt{|\phi
|}\epsilon_{IJK} e^I \wedge \Omega^{JK}(\omega) \label{action2},
\eq
where $e^I_a$ is the soldering one-form, $\omega^{IJ}_a$ is the spin
connection and $\Omega_{ab}^{IJ}$ is the curvature of $\omega_a^{IJ}$.
The quantum theory is determined by the partition functional
\bq
\mathcal{Z}=\int
\mathcal{D}[\phi]\mathcal{D}[e]\mathcal{D}[\omega]\mathbf{e}^{-i\int_{\mathcal{M}}\sqrt{|\phi
|}\epsilon_{IJK} e^I \wedge \Omega^{JK}} \label{Partition1}.
\eq
We introduce the triangulation $\triangle$ of $\mathcal{M}$ and its
dual $\triangle^*$ as the regulator
 of the system. Point, segment and triangle of $\triangle$ are denoted as $p, s$
 and $t$, and vertex, edge and face of $\triangle^*$ are denoted as $v, e$ and $f$.
 The triad field $e_a^I$ is smeared along segments, and the term
 $\sqrt{|\phi |}\Omega^{IJ}_{ab}$, as a $\mathfrak{su(2)}$ valued two
 form, is smeared along the faces of $\triangle^*$ that dual to the segments,
\ba
E^I_s=\int_s e^I,\qquad \Omega^{IJ}_f(\phi)=\int_f \sqrt{|\phi
|}\Omega^{IJ}.\label{smear}
\ea
The triangulation are supposed to
be fine enough, so that on the faces the scalar $\sqrt{|\phi(x)|}$ can
be approximated by a constant $\sqrt{|\Phi_f|}$. Then Eq.(\ref{smear}) reads
\bq
\Omega^{IJ}_f(\phi)=\int_f \sqrt{|\Phi_f|} \Omega^{IJ}=\sqrt{|\Phi_f|}\Omega^{IJ}_f,
\eq
where $\Omega^{IJ}_f=\int_f \Omega^{IJ}_{ab}$, and
$\sqrt{|\Phi_f|}$ can be treated as the weight of the variable
$\Omega^{IJ}_f$. The smeared curvature $\Omega^{IJ}_f$ can be
related to the holonomy $U_f$ along the boundary $e^1_f\circ \ldots
\circ e^n_f$ of $f$ as
\bq
g_{e^1_f}\circ g_{e^2_f}\circ \ldots\circ g_{e^n_f}=
U_f=e^{\Omega_f}=\mathbf{1}_g+\Omega_f+\ldots\label{Holonomy},
\eq
where $g_{e^i_f}=\mathcal{P}exp(\int_{e^i_f} \omega^{IJ})$ is the group
element associate to each edge. By the regularization, the degrees of freedom left
are $(E^I_s,g_e,\Phi_f)$, and the partition functional takes the form
\bq
\mathcal{Z}= \sum_{\triangle}\int\prod_s dE^I_s\prod_e dg_e
\prod_f d\Phi_f  \mathbf{e}^{-i\sum_s tr(E_s(\sqrt{|\Phi_f|} \Omega_f))}
\label{Partition2},
\eq
where the weight $\mathbf{w}(\triangle)$
associated to each triangulation are assumed to be the same and set
to $\mathbf{1}$ for simplicity. This regularized partition functional will approach
 Eq.(\ref{Partition1}) when the triangulation becomes
finer and finer. For a fixed triangulation, integrating out the
triad $E^I_s$, we have
\ba
\mathcal{Z}_{\triangle}
=\int\sum_{j_f}\prod_f d\Phi_f \prod_e dg_e \prod_f \triangle_{j_f}\chi^{j_f}(G_{e^1_f}\circ  \ldots \circ
G_{e^n_f}),\nonumber
\ea
where $G_{e^i_f}=\mathcal{P}exp(\int_{e^i_f}\sqrt{|\Phi_f|} \omega^{IJ})$ is a $SU(2)$
element that depends on $(\Phi_f, g_{e^i_f})$,
$\triangle_{j_f}$ is the dimension of the representation $j_f$,
 and $\chi^{j_f}(g)$ is the character of the group element $g$ in the representation $j_f$.
Though the vertex amplitude $\mathcal{A}_v$ is not worked out explicitly, the dynamics
does involve both the representation $j$ and the scalar $\phi$.
\begin{figure}
\includegraphics[scale=0.55]{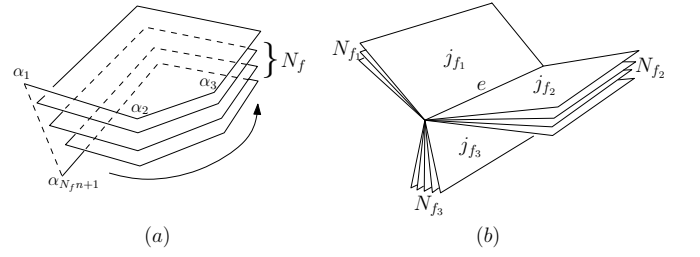}
\caption{(a) The $N_f$ ``virtual'' faces.
(b) Three ``bunch'' of faces that joint at each edge.} \label{Nface}
\end{figure}
Now we consider the sector that the weight $\sqrt{|\Phi_f|}$ is
discrete, i.e., the discretization of the scalar field. We rewrite
the action (\ref{action2}) as \bq
S[e,\omega,\phi]=\int_{\mathcal{M}}\mathcal{C}
\sqrt{\frac{|\phi|}{\mathcal{C}^2}} \epsilon_{IJK} e^I \wedge
\Omega^{JK}(\omega) \label{action3} \eq and let
$\sqrt{|\Phi_f|/\mathcal{C}^2}$ to take values in
$\mathbb{Z^+}\cup\{0\}$, which are denoted as $N_f$. The constant
multiplier $\mathcal{C}$ will not affect the physics, and the steps
of $\mathcal{C} N_f \epsilon_{IJK} e^I \wedge \Omega^{JK}(\omega)$
in the action can be made as small as possible by choosing
appropriate $\mathcal{C}$. This ensures that the regularized
partition functional (\ref{Partition2}) can be approximated as good
as one wants. The partition functional now reads \bq
\mathcal{Z}^N_{\triangle}=\int \prod_s dE^I_s \prod_e dg_e
\sum_{N_f} \rho(\overrightarrow{N}) \mathbf{e}^{-i\sum_s tr(E_s(N_f
\Omega_f))} \label{Partition4}, \eq where $\overrightarrow{N}$
denotes $\{N_{f_1},N_{f_2},\ldots\ldots\}$, and
$\rho(\overrightarrow{N})$ is the weight for each configuration
$\overrightarrow{N}$, which comes from the measure $d\Phi_f$ in
Eq.(\ref{Partition2}). Integrating out $E^I_s$ we have \ba
\mathcal{Z}^N_{\triangle} &=&\int\prod_e dg_e \sum_{N_f}
\rho(\overrightarrow{N}) \prod_f
 \delta(\overbrace{U_f\circ\ldots\circ U_f}^{{\scriptscriptstyle N_f}})\nonumber\\
&=&\sum_{N_f} \sum_{j_f} \rho(\overrightarrow{N})\int\prod_e dg_e  \prod_f\{\triangle_{j_f}
 \nonumber\\
&&\delta^{\alpha_{N_fn+1}}_{\alpha_1}(R^{j_f}(g_{e^1_f})^{\alpha_1}_{\alpha_2}R^{j_f}(g_{e^2_f})
^{\alpha_2}_{\alpha_3}\ldots R^{j_f}(g_{e^n_f})^{\alpha_n}_{\alpha_{n+1}})\nonumber\\
&&(R^{j_f}(g_{e^1_f})^{\alpha_{n+1}}_{\alpha_{n+2}}\ldots
 R^{j_f}(g_{e^n_f})^{\alpha_{2n}}_{\alpha_{2n+1}})\ldots\nonumber\\
&&(R^{j_f}(g_{e^1_f})^{\alpha_{(N_f-1)n+1}}_{\alpha_{(N_f-1)n+2}}\ldots
R^{j_f}(g_{e^n_f})^{\alpha_{N_fn}}_{\alpha_{N_fn+1}})\},\label{NfRe}
\ea where $R^j(g)^{\alpha}_{\beta}$ denotes the representation
matrix of $g$ which belongs to the representation $j$. We have
chosen a special but natural ordering of the $N_f$ copies of the
group elements $g_{e^i_f}$ in the above derivation, which plays a
key role in our model. The ambiguity caused by this ordering can be
controlled by means of fine-enough triangulations. The result is
interpreted as that for each face $f$ of $\triangle^*$ we associate
$N_f$ copies of ``virtual'' faces which are all colored by the same
representation $j_f$. The indexes of the representation matrixes of
edges which bound the $N_{f}$ faces are contracted with each other
following the order expressed in Eq.(\ref{NfRe}). See
FIG.\ref{Nface}(a) for a graphic presentation, where the contraction
wind around each face $f$ for $N_f$ times. There will be one group
element, denoted as $g_{e^n_f}$, for each face $f$ whose contraction
with the next one will cross the ``virtual'' faces. We will show
that  the choices of these \textit{crossing edges} do not affect the
physics. For each edge $e$ of $\triangle^*$, there are three
``bunches'' of faces which joint at it (see FIG.\ref{Nface}(b)).
According to Eq.(\ref{NfRe}), there is one following integral for
each edge, \ba &&\int dg_e \bigotimes^{N_{f_1}}_{i=1}
R^{j_{f_1}}(g_e)^{\alpha_i}_{\beta_i} \bigotimes^{N_{f_2}}_{j=1}
R^{j_{f_2}}(g_e)^{\gamma_j}_{\sigma_j}
 \bigotimes^{N_{f_3}}_{k=1} R^{j_{f_3}}(g_e)^{\rho_k}_{\lambda_k}=\sum_\iota \nonumber\\
&&\iota^{(\alpha_1\ldots \alpha_{N_{f_1}})(\gamma_1\ldots
  \gamma_{N_{f_2}})(\rho_1\ldots \rho_{N_{f_3}})}\iota^*_{(\beta_1\ldots
  \beta_{N_{f_1}})(\sigma_1\ldots \sigma_{N_{f_2}})(\lambda_1\ldots
  \lambda_{N_{f_3}})}\nonumber{}\\
&&=P^{\bigotimes_{N_{f_1}} j_{f_1}\bigotimes_{N_{f_2}}
j_{f_2}\bigotimes_{N_{f_3}} j_{f_3}}_{inv}\label{EdgeInt},
\ea
where $P^{ \bigotimes_{N_{f_1}} j_{f_1}\bigotimes_{N_{f_2}}
j_{f_2}\bigotimes_{N_{f_3}} j_{f_3}}_{inv}$ is the projector into
the invariant subspace $Inv$,
and $\iota^{\alpha_1\ldots \rho_{N_{f_3}}}$ form an orthonormal basis of the
invariant subspace. These tensors are symmetric with respect to the indexes from the same
``bunch'' of faces. For the sake of readability, we will neglect these parentheses in the following context.
  This operator $P_{inv}$ and the subspace $Inv$ are not null
only if the corresponding representations coupled together satisfy
certain compatible conditions \cite{Brink}.
\begin{figure}
 \includegraphics[scale=0.47]{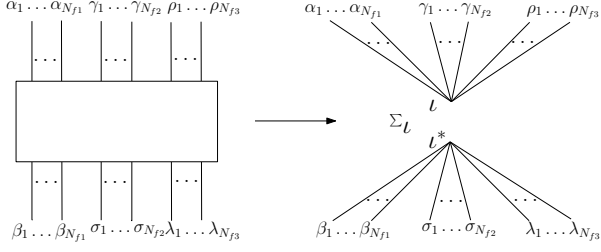}
\caption{The edge integral of Eq.(\ref{EdgeInt}).}
\label{edgeInt}
\end{figure}
\begin{figure}
 \includegraphics[scale=0.6]{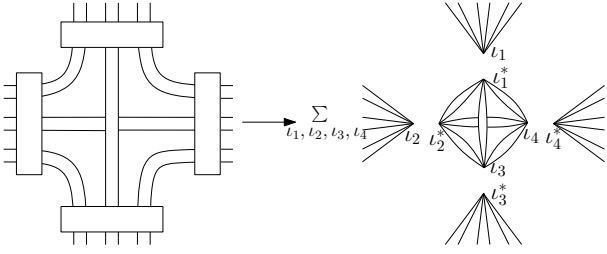}
\caption{The couplings at a vertex: the double lines represent the
  ``bunches'' of representations.}
\label{vertex}
\end{figure}
The edge integral (\ref{EdgeInt}) can be illustrated as
FIG.\ref{edgeInt} by means of the graphical techniques in performing
$SU(2)$ tensor calculus \cite{Penrose,Brink}, where the rectangle on
the left denotes the integral $\int_e dg_e$. According to the
structure of $\triangle^*$, there are four edges which joint at each
vertex, and hence six "bunches" of faces joint at each vertex. Thus,
for each vertex we have four integrals of the form (\ref{EdgeInt})
with the corresponding indexes contracted, i.e.,
\ba
&&\int dg_{e^1}dg_{e^2}dg_{e^3}dg_{e^4}\nonumber{}\\
&&\{\bigotimes^{N_{f_1}}_{i=1}
R^{j_{f_1}}(g_{e^1})^{\alpha^1_i}_{\lambda^1_i} \bigotimes^{N_{f_2}}_{j=1}
R^{j_{f_2}}(g_{e^1})^{\alpha^2_j}_{\lambda^2_j}\bigotimes^{N_{f_3}}_{k=1}
R^{j_{f_3}}(g_{e^1})^{\alpha^3_k}_{\lambda^3_k}\nonumber\\
&& \bigotimes^{N_{f_1}}_{l=1}
R^{j_{f_1}}(g_{e^2})^{\lambda^1_l}_{\beta^1_l} \bigotimes^{N_{f_4}}_{m=1}
R^{j_{f_4}}(g_{e^2})^{\alpha^4_m}_{\lambda^4_m} \bigotimes^{N_{f_5}}_{n=1}
R^{j_{f_5}}(g_{e^2})^{\alpha^5_n}_{\lambda^5_n}\nonumber\\
&&\bigotimes^{N_{f_5}}_{p=1}
R^{j_{f_5}}(g_{e^3})^{\lambda^5_p}_{\beta^5_p} \bigotimes^{N_{f_2}}_{q=1}
R^{j_{f_2}}(g_{e^3})^{\lambda^2_q}_{\beta^2_q} \bigotimes^{N_{f_6}}_{r=1}
R^{j_{f_4}}(g_{e^3})^{\alpha^6_m}_{\lambda^6_m} \nonumber\\
&&\bigotimes^{N_{f_6}}_{s=1}
R^{j_{f_6}}(g_{e^4})^{\lambda^6_s}_{\beta^6_s} \bigotimes^{N_{f_4}}_{t=1}
R^{j_{f_4}}(g_{e^4})^{\lambda^4_t}_{\beta^4_t} \bigotimes^{N_{f_3}}_{w=1}
R^{j_{f_3}}(g_{e^4})^{\lambda^3_w}_{\beta^3_w}\}\nonumber\\
&&=\sum_{\iota_{(1)}\iota_{(2)}\iota_{(3)}\iota_{(4)}}\{
\iota_{(1)}^{\alpha^1_1\ldots\alpha^1_{N_{f1}}\alpha^2_1\ldots\alpha^2_{N_{f2}}
\alpha^3_1\ldots\alpha^3_{N_{f3}}}\nonumber\\
&&\iota^{\alpha^4_1\ldots\alpha^4_{N_{f4}}\alpha^5_1\ldots\alpha^5_{N_{f5}}}_{(2)
\quad\quad\quad\quad\quad\enspace1\beta^1_1\ldots\beta^1_{N_{f1}}}
\iota^{*\alpha^6_1\ldots\alpha^6_{N_{f6}}}_{(3)\quad\quad\enspace\beta^5_1\ldots\beta^5_{N_{f5}}
\beta^2_1\ldots\beta^2_{N_{f2}}}\nonumber\\
&&\iota^*_{(4)\beta^6_1\ldots\beta^6_{N_{f6}}\beta^4_1\ldots\beta^4_{N_{f4}}
\beta^3_1\ldots\beta^3_{N_{f3}}}\mathcal{A}_v(\iota_{(1)},\iota_{(2)},\iota_{(3)},\iota_{(4)}
)\}\nonumber, \ea where the free indexes will contract with that of
the nearby vertices. The resulted amplitude reads \ba
&&\mathcal{A}_v(\iota_{(1)},\iota_{(2)},\iota_{(3)},\iota_{(4)})\nonumber{}
=\begin{array}{c}\includegraphics[scale=0.27]{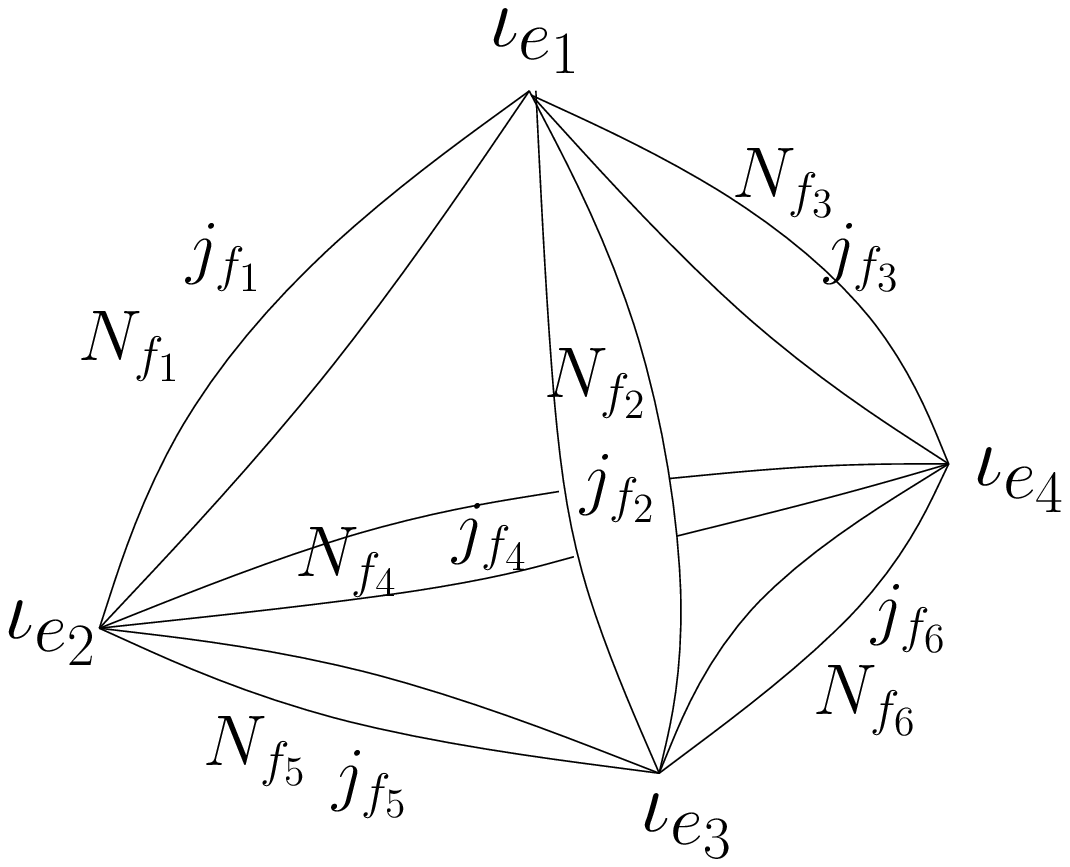}\end{array}\\
&=&\iota^*_{(1)\lambda^1_1\ldots\lambda^1_{N_{f1}}\lambda^2_1\ldots
\lambda^2_{N_{f2}}\lambda^3_1\ldots\lambda^3_{N_{f3}}}\iota^
{*\lambda^1_1\ldots\lambda^1_{N_{f1}}}_{(2)\quad\enspace\enspace
\lambda^4_1\ldots\lambda^4_{N_{f4}}\lambda^5_1\ldots\lambda^5_{N_{f5}}}\nonumber{}\\
&&\iota_{(3)}^{\lambda^5_1\ldots\lambda^5_{N_{f5}}\lambda^2_1\ldots
\lambda^2_{N_{f2}}\lambda^6_1\ldots\lambda^6_{N_{f6}}}\iota^
{\lambda^3_1\ldots\lambda^3_{N_{f3}}\lambda^4_1\ldots\lambda^4_{N_{f4}}}
_{(4)\quad\quad\quad\enspace\enspace\enspace\lambda^6_1\ldots\lambda^6_{N_{f6}}}\label{Av},
\ea
where the Einstein summation convention is adopted.
 The result is represented by means of the graphical technics as FIG.\ref{vertex}.
The presence of crossing edges only cause a permutation of the
indexes of the same ``bunch'' of faces and hence does not affect the
amplitudes $A_v$. Thus we arrived at the final form of the partition
functional of Eq.(\ref{Partition4}),
\bq
\mathcal{Z}^N=\sum_{\triangle}\sum_{{\scriptscriptstyle
(j_f,N_{f},\iota_e)}}\rho(\overrightarrow{N})
\prod_{{\scriptscriptstyle f\in\triangle^*}}
\triangle_{j_f}\prod_{{\scriptscriptstyle
v\in\triangle^*}}\mathcal{A}_v(j_f,N_f,\iota_e)\nonumber{}.
\eq

\begin{figure}
\includegraphics[scale=0.49]{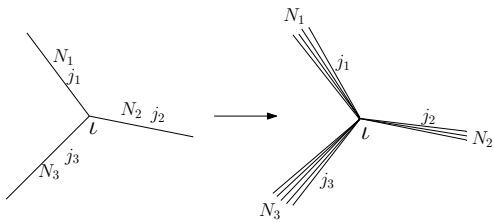}
\caption{Casting $\hat{T}_{\gamma,N,j,\iota}$ to $T_{\gamma,N,j,\iota}$.}
\label{network}
\end{figure}

The boundary states that encode the physical information on a
hypersurface $\Sigma$ which intersects with the spinfoams can be
described as the ``generalized'' $SU(2)$ spin-network states
$\tilde{T}_{\gamma,N,j,\iota}$, consisting of network-like graphs
$\gamma$ with nodes labeled by the intertwinors $\iota_e$ and links
labeled by both the representations $j_f$ and the integers $N_f$.
The remarkable result comes up as that these states
$\tilde{T}_{\gamma,N,j,\iota}$ can be casted into the pure $SU(2)$
spin-network states $T_{\gamma,N,j,\iota}$, which constitute the
basis of the kinematical Hilbert space $\mathcal{H}_{kin}$ of 3D
quantum gravity \cite{AshtekarReview,RovelliBook,ThiemannBook}. This
is done by replacing the $(N_f,j_f)$-labeled links by $N_f$ copies
of $j_f$-labeled links which joint at the same nodes as
FIG.\ref{network}. With the different choices of $N_f$ and $j_f$,
$T_{\gamma,N,j,\iota}$ will run over a special subset $\mathbb{T}$ of $\mathcal{H}_{kin}$.
Hence the information of the scalar field on the
hypersurface are totally encoded into the standard $SU(2)$ spin-network
states $T_{\gamma,N,j,\iota}$ or the states of quantum geometry. Thus
the kinematics of this model has been casted into that of
the 3D quantum geometry.
\begin{figure}
\includegraphics[scale=0.48]{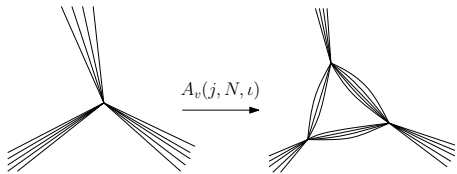}
\caption{Evolutions of the nodes.}
\label{trans}
\end{figure}
The evolution of $T_{\gamma,N,j,\iota}$ is generated by the vertices
as FIG.\ref{trans}, and the transition amplitudes
$\mathcal{A}_{v}(j_f,N_f,\iota_e)$ can be calculated in a purely
combinational way by means of the recouplings of angular momentums.
The dynamics can be interpreted as the quantum evolutions from the
initial boundary states $T^i_{\gamma,N,j,\iota}$ to the final
boundary states $T^f_{\gamma,N,j,\iota}$. This is equivalent to the
choice of the physical inner product $<T^f,T^i>_{phy}$ between the
kinematical states. By the formal relation \cite{RovelliProjector}
\bq
<T^f,T^i>_{phy}=\int D[N] <T^f,\mathbf{e}^{-i\int_{\Sigma}
\hat{\mathbf{H}}[N]}T^i>_{kin}\nonumber,
\eq
one may go inversely to
construct the Hamiltonian operator $\hat{\mathbf{H}}$ defined on
$\mathcal{H}_{kin}$ which generates this dynamics.
 The new  dynamics will reduce to that of Ponzano-Regge model,
as one would expect, when $\phi$ is assumed to be a constant
 field and $N_f$ is set to 1 for all faces.
It is a natural extension of the dynamics of pure
gravity by the new transition amplitudes (\ref{Av}) coming also from the couplings
of $SU(2)$ representations.
As a generalization of Penrose's idea \cite{Penrose}, both the kinematics and
the dynamics of the scalar matter and gravity are built into these couplings of the angular momentums.
In the full quantum situation,
the physics of this system can be casted into the dynamics of the pure
 quantum geometry, while,
in the semiclassical situation, the physics manifests itself as the
dynamics of the geometry coupled with the scalar matter. Thus the
scalar field can be viewed as a phenomena emerged from the
microscopical system of quantum spinfoams. This gives rise to a
remarkable and radical observations that the geometry and matter
fields appearing in the low energy scale may originate from a single
microscopical construction, which may be spinfoams in accordance
with above viewpoint.

To summarize, the spinfoam model of the system (\ref{action2}) is
worked out. This model shed some new lights on the issue of matter
couplings in LQG and spinfoam formalism. The distinct property of
this model gives rise to a radical observation on the issue of
unification of geometry and matter. Being the convergent point of
distinct approaches of quantum gravity, spinfoams may also be the
convergent point of geometry and matter.

The authors would like to thank Dah-Wei Chiou and Muxin Han for
discussions. This work is a part of project 10675019 supported by
NSFC.

\newcommand\AL[3]{~Astron. Lett.{\bf ~#1}, #2~ (#3)}
\newcommand\AP[3]{~Astropart. Phys.{\bf ~#1}, #2~ (#3)}
\newcommand\AJ[3]{~Astron. J.{\bf ~#1}, #2~(#3)}
\newcommand\APJ[3]{~Astrophys. J.{\bf ~#1}, #2~ (#3)}
\newcommand\APJL[3]{~Astrophys. J. Lett. {\bf ~#1}, L#2~(#3)}
\newcommand\APJS[3]{~Astrophys. J. Suppl. Ser.{\bf ~#1}, #2~(#3)}
\newcommand\JCAP[3]{~JCAP. {\bf ~#1}, #2~ (#3)}
\newcommand\LRR[3]{~Living Rev. Relativity. {\bf ~#1}, #2~ (#3)}
\newcommand\MNRAS[3]{~Mon. Not. R. Astron. Soc.{\bf ~#1}, #2~(#3)}
\newcommand\MNRASL[3]{~Mon. Not. R. Astron. Soc.{\bf ~#1}, L#2~(#3)}
\newcommand\NPB[3]{~Nucl. Phys. B{\bf ~#1}, #2~(#3)}
\newcommand\PLB[3]{~Phys. Lett. B{\bf ~#1}, #2~(#3)}
\newcommand\PRL[3]{~Phys. Rev. Lett.{\bf ~#1}, #2~(#3)}
\newcommand\PR[3]{~Phys. Rep.{\bf ~#1}, #2~(#3)}
\newcommand\PRD[3]{~Phys. Rev. D{\bf ~#1}, #2~(#3)}
\newcommand\SJNP[3]{~Sov. J. Nucl. Phys.{\bf ~#1}, #2~(#3)}
\newcommand\ZPC[3]{~Z. Phys. C{\bf ~#1}, #2~(#3)}
\newcommand\CQG[3]{~Class. Quant. Grav. {\bf ~#1}, #2~(#3)}

\end{document}